\documentstyle[11pt,aaspp4,flushrt]{article} 

\input psfig

\newcommand{\beq}{\begin{equation}}
\newcommand{\eeq}{\end{equation}}

\def\lsim{\mathrel{\mathpalette\@versim<}}
\def\gsim{\mathrel{\mathpalette\@versim>}}

\def\kpa{k_{\parallel}}

\begin{document}
\title{The Proton Distribution Function in Weakly Magnetized Turbulent
Plasmas} \author{Andrei Gruzinov$^{\dagger}$ \& Eliot
Quataert$^{\dagger \dagger}$} \affil{$^{\dagger }$ Institute for
Advanced Study, School of Natural Sciences, Princeton, NJ 08540}
\affil{$^{\dagger \dagger }$ Harvard-Smithsonian Center for
Astrophysics, 60 Garden Street, Cambridge, MA 02138}

\begin{abstract}

We calculate the proton distribution function due to heating by
subsonic (Alfvenic) turbulence in a weakly magnetized collisionless
plasma.  The distribution function is nonthermal.  For
non-relativistic energies, it is an exponential of the
magnitude of the proton velocity.  For ultra-relativistic energies, it
can be characterized as a power law with a momentum-dependent slope.

\end{abstract}

\section{Introduction}
The proton distribution function (PDF) in a collisionless plasma
depends on the history of the plasma. If, for example, the plasma is
heated by shocks, the PDF is believed to be a power law at high
energies, while for the thermal particles, the form of the PDF is not
thought to be universal (e.g., Blandford \& Eichler 1987).  In this
paper we calculate the PDF for a weakly magnetized plasma (the Alfven
speed $v_A$ is less than the ion thermal speed $v_i$) heated by
subsonic (Alfvenic) turbulence.

Although we do not have a fully quantitative theory for the
turbulence (\S2), we are still able to calculate the distribution function (\S3). This is due to a special feature of particle
heating by low frequency, subsonic, waves (\S3.1). In \S4 we mention a
possible application of our results to accreting black holes. We
summarize our results in \S5.

\section{Proton Heating by Subsonic Turbulence}

The nature of subsonic turbulence in a collisionless weakly magnetized
plasma is not fully understood. A plausible scenario, based on
Goldreich and Sridhar (1995), is as follows. Large scale motions
(larger than the ion Larmor radius $\rho$) are nearly dissipationless
and incompressible. If the characteristic velocity of the energy
containing eddies (eddies of the largest scale $L$) is small,
$V\lesssim v_A\ll v_i$, the dynamics is adequately described by
reduced MHD (incompressible MHD without slow modes, e.g. Strauss 1976,
Ng \& Bhattacharjee 1996).  As in hydrodynamics, the turbulent energy
cascades to small scales where it is dissipated. The cascade is
described by a Kolmogorov 5/3 law, but it is strongly anisotropic,
with most of the energy residing in quasi-perpendicular perturbations. Typical wavevectors ${\bf k}$ are nearly perpendicular to the local magnetic field, so that perturbations are strongly elongated along the local magnetic field.  The energy per unit mass
in eddies of perpendicular wavenumber $\sim k_{\perp}$ is $\sim
\epsilon ^{2/3}k_{\perp}^{-2/3}$, where $\epsilon \sim V^3/L$ is the
net energy dissipation rate per unit mass. The degree of anisotropy is
determined by the requirement that the Alfven frequency not exceed the
nonlinear frequency, which yields $k_{\parallel} \lesssim
k_{\perp}^{2/3} L^{-1/3}$.

``Viscous'' heating occurs when the perpendicular size of the eddy is
$\sim \rho$ (Quataert 1998, Gruzinov 1998, Quataert \& Gruzinov
1998). The small-scale eddies are essentially Alfven waves. Alfven
waves with a perpendicular wavenumber $\sim \rho ^{-1}$ have a
non-zero parallel magnetic field perturbation (in contrast to long
wavelength perturbations).  As a result, particles interact with waves
through ``magnetic mirror'' forces, i.e., through the coupling of the
magnetic moment of the particle and the parallel gradient of the
magnetic field. A charged particle moving along the field will be
randomly accelerated and decelerated along the field, and will
therefore diffuse in parallel velocity space.

Strong parallel diffusion requires $|v_{\parallel}| \lesssim v_A$,
where $v_{\parallel}$ is the component of the particle's velocity
along the magnetic field.  This can be understood as follows.  A
parallel magnetic field perturbation of frequency $\omega$ and
parallel wavenumber $k_{\parallel}$ will accelerate particles that
satisfy the resonance condition
$v_{\parallel}=\omega/k_{\parallel}$. If the turbulence were strictly
a superposition of linear Alfven waves, the resonance condition would
imply that only particles with $|v_{\parallel}| = v_A$ are accelerated
(because Alfven waves have the dispersion relation $\omega =
|k_{\parallel}| v_A$).

It is likely, however, that the strong turbulence of interest is not
fully describable as linear Alfven waves.  In this case, at the
smallest scales ($k_{\perp}\sim \rho ^{-1}$), perturbations of all
frequencies $\omega \lesssim \omega _{nl}$ and wavenumbers
$k_{\parallel}\lesssim \omega _{nl }/v_A$ will exist, where $\omega
_{nl} \sim \epsilon ^{1/3}\rho ^{-2/3}$ is the nonlinear frequency on
scales $\sim \rho$.  The quantity $\omega/k_{\parallel}$ then takes on
values between 0 and $\infty$. This does not, however, mean that a
particle with an arbitrary parallel velocity will be
accelerated. Acceleration of high parallel-velocity particles,
$|v_{\parallel}| > v_A$, will be insignificant because perturbations
resonating with a high parallel-velocity particle must have small
$k_{\parallel}$. These perturbations will exert negligibly small
magnetic mirror forces for two reasons.  The parallel force is
proportional to the parallel gradient of the parallel magnetic field
perturbation (see eq. [\ref{eom}]).  In addition, the parallel
magnetic field perturbation is itself proportional to the parallel
gradient of the electric field perturbation, because an incompressible
two dimensional plasma flow with $\kpa =0$ does not perturb the
parallel magnetic field.\footnote{Formally, this follows from the
small $\kpa$ limit of the plasma dielectric tensor.}  Magnetic mirror
forces are therefore $\sim \kpa^2$; perturbations with small
$k_{\parallel}$ are consequently negligible.

\section{The Evolution of the Proton Distribution Function}

Diffusion in parallel velocity due to turbulent heating can be
described by the equation
\begin{equation}
{\partial f \over \partial t} ={\partial \over \partial p_{\parallel}} \left(
D_{\parallel} {\partial \over \partial p_{\parallel}} f\right),
\end{equation}
where $D_{\parallel}$ is the parallel momentum
diffusivity. Anisotropies in momentum space are quickly erased by
high-frequency small-scale electromagnetic instabilities. Parallel
momentum diffusion is then equivalent to an isotropic diffusion in
momentum space,
\begin{equation}
{\partial f \over \partial t}={1\over p^2} {\partial \over \partial p}
\left( p^2 D {\partial \over \partial p} f\right), \label{diff}
\end{equation}
where \beq D = {1 \over 2} \int^{1}_{-1} d \mu \mu^2 D_\parallel
\label{D} \eeq and $\mu$ is the cosine of the particle's pitch angle
(the angle between the velocity vector and the local magnetic field).

Suppose that protons are strongly heated, so that their final thermal
energy is much larger than their initial thermal energy.  In this
case, the final distribution function depends primarily on the
properties of the diffusion under consideration, but only weakly on
the initial distribution function.  If $D$ does not depend on $p$, the
final distribution function will be Maxwellian.\footnote{The Green's
function of equation (\ref{diff}) with $D={\rm const.}$ is a
Gaussian.}  As we show below, for our problem $D \propto \gamma p$,
where $\gamma $ is the particle's Lorentz factor.  Since $D$ is not
constant, diffusion establishes a non-Maxwellian PDF.

\subsection{The Diffusion Coefficient}

We first calculate the parallel diffusivity, $D_{\parallel}$.  The
parallel motion of a particle interacting with low frequency
perturbations is described by
\begin{equation}
{dv_{\parallel}\over dt} =-{v_{\perp}^2\over 2}{\nabla _{\parallel} B\over B},
\label{eom} \end{equation}
where we have assumed $|v_{\parallel}| \ll v_{\perp}$. Since
$p_{\parallel} = \gamma mv_{\parallel}$, the parallel momentum
diffusivity is
\begin{equation}
D_{\parallel} \propto \gamma ^2v_{\perp}^4\Phi (v_{\parallel}/ v_A).
\end{equation}
$\Phi (x)$ is a positive dimensionless function, which is $\sim 1$ for
$|x|\lesssim 1$, and rapidly approaches zero at larger $|x|$.  This is
because only particles with sub-Alfvenic parallel velocities efficiently
interact with the turbulence (see \S2).  $\Phi(x)$ cannot be
calculated in detail because the spectrum of turbulence at small
scales, where nonlinear and kinetic effects are both important, is
unknown.\footnote{For linear MHD Alfven waves, $\Phi(x) \propto
\delta(|x| - 1)$.}  Because we are interested in $v_A \ll v_i$, however,
our final answer will be independent of $\Phi$.

The isotropic diffusivity, $D$, can be obtained from $D_\parallel$
using equation (\ref{D}), which yields 
\begin{equation}
D\propto \gamma ^2v^4\int d \mu \mu^2 (1 - \mu^2)^2 \Phi (\mu v/v_A)
\propto \gamma ^2 v. \label{Df}
\end{equation}

\subsection{The Distribution Function}

For non-relativistic protons, $D\propto p$. Solving equation
(\ref{diff}) for an initial PDF $f_0(p) \propto \delta(p)$ gives
\begin{equation}
f(p)= {1 \over \pi (Tm)^{3/2}} \exp (-2p/\sqrt{Tm} ),
\end{equation}
where the ``temperature'' is defined by the usual expression
$<p^2>=3Tm$.  Although the PDF is nonthermal, the number of
suprathermal protons is exponentially small.  Subsonic turbulence is
therefore intrinsically inefficient at accelerating particles in the
non-relativistic limit.

At ultra-relativistic energies, $D\propto p^2$.  Solving equation
(\ref{diff}) for $f_0(p) \propto \delta(p - p_i)$, gives
\begin{equation}
f(p) \propto \left( {p\over p_f}\right)^{-\alpha (p)},
\end{equation}
where
\begin{equation}
\alpha (p)={7\over 2}+{\log (p/p_f)\over \log (p_f/p_i)}
\end{equation}
and $p_f$ is the mean momentum of the final PDF. At
relativistic energies, there is a significant population of particles
above the mean energy of the plasma.  Realizing this efficient
acceleration, however, requires that the bulk of the plasma be heated
to at least mildly relativistic energies, which limits its practical
significance.

Using equation (\ref{diff}) and equation (\ref{Df}), we numerically
calculated the distribution function for a plasma that was heated from
non-relativistic to mildly relativistic energies.  Figure 1 shows the
calculated distribution in energy space, $dN/dE$, where $E=(\gamma
-1)mc^2$.  Also shown is a Maxwellian with a temperature of $0.2{\rm
GeV}$, which has the same mean energy per particle. The viscously
heated PDF has a noticeable nonthermal tail of relativistic particles.
 
\section{Application to Accretion Flows}

A possible application of our results is to hot, quasi-spherical,
collisionless plasmas which have been proposed to exist near accreting
black holes.  By spherical accretion we mean any of the proposed
accretion models, Bondi (Bondi 1952), ion tori (Rees et al 1982) or
advection-dominated accretion flows (Narayan \& Yi 1995). Our calculated PDF may 
be relevant to these gravitationally confined
plasmas (which must have $v_A \lesssim v_i$ and, as we have argued
elsewhere, are likely to have $v_A \ll v_i$).  For protons with
roughly virial energies near the Schwarzschild radius of a black hole,
the PDF calculated in the previous section is noticeably
non-Maxwellian, with an excess of high-energy protons.

In such hot plasmas, even a Maxwellian PDF leads to interesting levels
of gamma-ray emission due to pion production in proton-proton
collisions (Mahadevan et al 1998; Ginzburg \& Syrovatskii 1964 give a
clear description of how to estimate the rates).  Our non-Maxwellian
PDF should produce distinct observable features in the gamma-ray
spectrum.  This will be calculated in a separate paper since
theoretically estimating the PDF for the purposes of gamma-ray
emission entails (in addition to the present analysis) a number of
additional complications, e.g., adiabatic heating of particles as well
as the possibility that the flow may capture and compress ambient
cosmic rays (Meszaros 1975) or pre-accelerated protons from a
Bondi-Hoyle bow shock.

\section{Summary}

We have calculated the proton distribution function due to heating by
subsonic (Alfvenic) turbulence in a weakly magnetized collisionless
plasma.  For non-relativistic energies, the PDF is nonthermal only to
the extent that it cuts off exponentially, rather than as a Gaussian,
with particle momenta.  By contrast, at mildly relativistic or
relativistic energies, a potentially observable ``suprathermal tail''
develops.

Our results seem to be relatively robust.  For plasmas heated to well
above their initial thermal energy, the distribution function is
determined solely by the momentum space diffusivity.  The diffusivity
we have calculated (eq. [\ref{Df}]) should be valid on quite general
grounds, provided the protons are indeed predominantly heated by
low-frequency, subsonic, waves.

\acknowledgements We thank John Bahcall, Ramesh Narayan, and Martin
Rees for useful discussions. AG was supported by NSF PHY-9513835.  EQ
was supported by an NSF Graduate Research Fellowship and by NSF Grant
AST 9423209.

\begin{figure}[htb]
\plotone{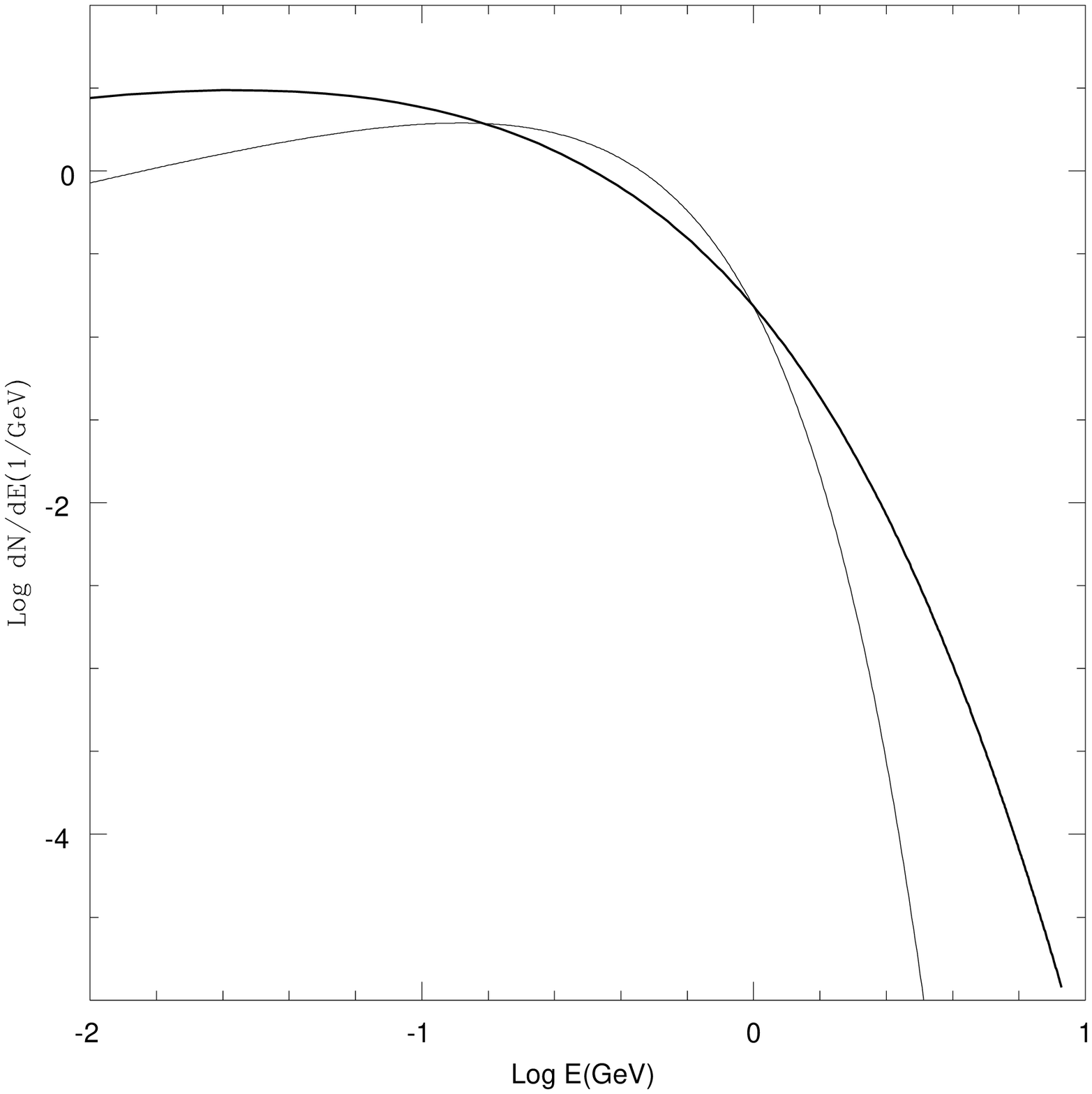}
\caption{Theoretically calculated proton distribution function due to
heating by subsonic turbulence (thick line). A Maxwellian with the
same mean energy per particle (thin line; $T = 0.2$GeV).}
\end{figure}

\end{document}